# Extraction of a single photon from an optical pulse

Serge Rosenblum[1]*, Orel Bechler[1]*, Itay Shomroni[1], Yulia Lovsky[1], Gabriel Guendelman[1] & Barak Dayan[1]

**Removing a single photon from a pulse is one of the most elementary operations that can be performed on light, having both fundamental significance[1,2] and practical applications in quantum communication[3-9] and computation[10]. So far, photon subtraction, in which the removed photon is detected and therefore irreversibly lost, has been implemented in a probabilistic manner with inherently low success rates using low-reflectivity beam splitters[1]. Here we demonstrate a scheme for the deterministic extraction of a single photon from an incoming pulse. The removed photon is diverted to a different mode, enabling its use for other purposes, such as a photon number-splitting attack on quantum key distribution protocols[11]. Our implementation makes use of single-photon Raman interaction (SPRINT)[12,13] with a single atom near a nanofibre-coupled microresonator. The single-photon extraction probability in our current realization is limited mostly by linear loss, yet probabilities close to unity should be attainable with realistic experimental parameters[13].**

[1] AMOS and Department of Chemical Physics, Weizmann Institute of Science, Rehovot 76100, Israel.
* These authors contributed equally to this work.

Photon subtraction has emerged not only as a test bed for investigating the foundations of quantum physics[1,2], but also as an essential instrument in the quantum information toolbox[3-10,14]. It is typically implemented in a heralded manner by using a low-reflectivity beam splitter directing a small fraction of the incoming light toward a detector[1]. In the rare occasions in which a detection occurs, one can infer that a single photon has been removed from the transmitted beam. The need to suppress the probability of more than one reflection leads to inherently low success rates. This procedure faithfully mimics the effect of the photon annihilation operator $\hat{a} = \sum_{n>0} \sqrt{n}|n-1\rangle\langle n|$ (where $|n\rangle$ is an $n$-photon Fock state), which is also probabilistic in nature, having higher success probabilities for a larger number of input photons. Accordingly, the annihilation operator conveys information on the number of photons in the incoming state. For example, for the classically common super-Poissonian distributions, the annihilation of one photon counterintuitively leaves the state with a higher mean photon number than before[1].

In recent years there has been a growing interest in the realization of deterministic single-photon extraction, in which exactly one photon is always removed from an incoming beam[15-19]. The corresponding operator $\hat{s} = |0\rangle\langle 0| + \sum_{n>0}|n-1\rangle\langle n|$, does not convey any additional information on the initial photon number, beyond the presence or absence of at least one photon in the incoming state. Indeed, if the probability of vacuum is negligible, the photon-number probability distribution of the output state is identical to that of the incoming one, only shifted down by one photon.

Here we experimentally demonstrate a scheme for the deterministic extraction of one photon from an incoming pulse, and its diversion to a different mode. The scheme makes use of the recently demonstrated[12] SPRINT mechanism, which was first proposed in ref. 20 and further studied in a series of theoretical works[17,18,21,22]. SPRINT occurs in an atomic (or atom-like) three-level Λ-system (see Fig. 1), in which each transition is coupled to a single direction of a single-mode waveguide with equal cooperativity $C \gg 1$ (where $2C=\Gamma/\gamma$, namely the ratio between the cavity-enhanced emission rate into one direction of the waveguide, and that into free space).

Assuming the atom is initiated in ground state $|\alpha\rangle$, which is coupled only to mode $\hat{a}$, an incoming long single-photon pulse in this mode (long compared to $1/\Gamma$) will eventually lead to a steady-state in which the output field of $\hat{a}$ drops to zero due to destructive interference between the incoming pulse and the re-emission from the atom. This leaves reflection to mode $\hat{b}$, and the associated Raman transfer of the atom to ground state $|\beta\rangle$ (to which $\hat{b}$ is coupled), as the only possible outcome.

Surprisingly, this destructive interference is sustained even for an input pulse containing many indistinguishable photons, as long as the photon flux does not exceed $\Gamma$[17,18]. The result is therefore a system in which transmission is blocked until a photon is reflected (and therefore extracted from the pulse), at which point it becomes transparent to the remaining photons, since in $|\beta\rangle$ the atom is decoupled from mode $\hat{a}$.

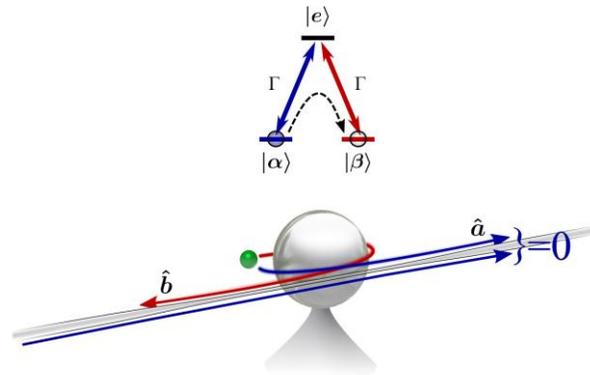

**Figure 1. Schematic depiction of single-photon Raman interaction (SPRINT).**

The two transitions in a three-level Λ-system (a single atom in this case) are coupled via a microresonator to different directions of a waveguide. A photon coming from the left is deterministically reflected (red arrows) due to destructive interference in the transmission (blue arrows), resulting in the Raman transfer of the atom from ground state $|\alpha\rangle$ to $|\beta\rangle$. The atom then becomes transparent to subsequent photons, which are therefore transmitted.

This mechanism, as well as any other mechanism that deterministically extracts the *first* photon arriving in a pulse, is in fact a temporally multimode mechanism and as such cannot be fully described by the single-mode operator $\hat{s}$ mentioned above. Specifically, the timing of the reflected photon contains information (a lower bound) on the timing of the transmitted photons, resulting in temporal entanglement between the reflected and transmitted pulses. Accordingly, the purity of the reflected and transmitted states (given by *n/(2n-1)*) gradually decreases from unity for a single photon input, approaching a lower bound of 1/2 at classically-high input photon numbers[23]. The purity can be retrieved by temporal postselection or spectral filtering at the price of reduced efficiency[23] (~69% for $n \gg 1$). However, there are many applications of deterministic single-photon extraction that do not require pure states, such as photon-number resolving detectors[15], as well as some protocols of quantum cryptography[24] and quantum information processing[25].

Our experimental realization is based on a single $^{87}$Rb atom coupled to an on-chip ultrahigh-Q microsphere resonator, into which light is evanescently coupled by a nanofibre. A transverse magnetic (TM) whispering gallery mode (WGM) of the microsphere is brought to resonance with the *F=1* → *F'=0* transition of the D$_2$ line of $^{87}$Rb by temperature tuning. The source of our atoms is a cloud of ~30·10$^6$ ultracold atoms released from 7 mm above the chip. A series of pulses in the nanofibre detects the presence of a single atom within the evanescent field of the WGM before the measurement, and verifies it afterwards (see Methods). The evanescent field of such modes has the property that clockwise (counterclockwise) propagating photons are predominantly $\sigma^+$ ($\sigma^-$)-polarized, and hence mostly drive $\Delta m_F = +1$ $(-1)$ transitions[26,27,28]. The probability of a photon having an undesired polarization is only ~6% (~4% overlap with the undesired circular polarization, and ~2% overlap with π-polarization, see Methods). Since, in addition, all *F=1* → *F'=0* transitions have equal oscillator strengths, this realization closely approximates the three-level Λ-system required for SPRINT. The average coherent coupling rate between this atomic transition and the TM mode was $g \sim 24$ MHz, significantly above the free-space amplitude decay rate of $\gamma = 3$ MHz. To optimize the SPRINT efficiency, the coupling rate between the cavity and the nanofibre was tuned to $\kappa_{ex} = 40$ MHz (see ref. 13), which together with the intrinsic cavity losses of $\kappa_i = 6.6$ MHz results in 48% linear loss in the absence of an atom. These parameters enable efficient interaction between incoming photons and the atom, as the total emission rate of the atom into both directions of the waveguide is larger than $\gamma$ by $4C = \frac{2g^2}{(\kappa_i + \kappa_{ex})\gamma} = 8.2 \gg 1$.

To experimentally demonstrate photon extraction, we sent 85 ns wide pulses with average photon numbers ranging from 0.2 to 11 onto the cavity-enhanced atom, which was initialized in $|\alpha\rangle$. Multiple single-photon detectors (five in each direction) acquired the photon statistics of the transmitted and the reflected light.

Figure 2a displays the mean number of reflected and transmitted photons as a function of the mean input photon number, both with and without the presence of an atom. The mean reflection, which was zero without atom, quickly saturates to ~1, corresponding to the fact that a reflected photon results in the passage of the atom to $|\beta\rangle$ where it is transparent to additional photons. For very low average photon numbers ($\bar{n}_{in} \ll 1$), the measured extraction efficiency is 40%. This value should be compared with the 52% extraction efficiency expected from an ideal photon extractor in the presence of 48% linear loss of the cavity. However, in cases where the loss occurred prior to the Raman passage of the atom to $|\beta\rangle$, the atom has a renewed chance of successfully extracting the next photon, if present. As a result, for large input photon numbers, the number of reflected photons is expected to approach ~0.73, ultimately limited by the probability of losing the extracted photon after the Raman passage. In our current system, however, every photon still has a

~4% overlap with a polarization that can drive the atom even in $|\beta\rangle$, possibly resulting in an additional reflection. This results in a slight upward slope in the reflection after saturation, leading to ~1 photon being reflected for 11 input photons. The transmission, in accordance, follows the expected complementary behavior of a single-photon subtracted pulse.

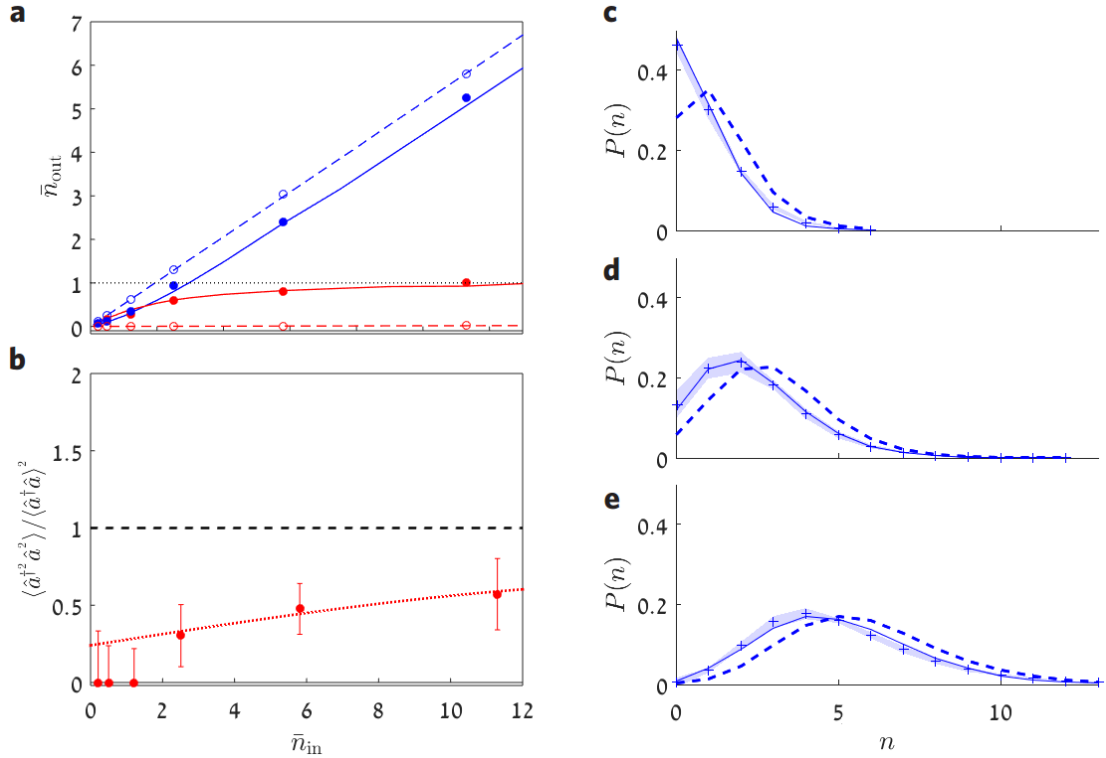

**Figure 2. Experimental results of single-photon extraction. a,** Measured mean number of reflected (red) and transmitted (blue) photons, with (solid markers) and without (empty markers) atom, as a function of the average number of input photons. The lines present the results of simulations with (solid) and without (dashed) atom, with no free parameters. Error bars are of equal size or smaller than the markers. The measurements closely follow the behavior expected from an ideal deterministic single-photon extractor in the presence of the measured 48% linear loss. **b,** Sub-Poissonian statistics of the reflected pulse (limited mostly by polarization impurity), demonstrating that the system preferentially extracts only a single photon, even if illuminated by more than ten photons. The dotted red line shows the result of simulations without free parameters. **c-e,** Reconstructed photon-number distributions for the transmitted light without (dashed lines) and with (crosses) atom, for 2.5, 5.8, and 11.3 photons at the input (corresponding to the three strongest-input data points in **a** and **b**). In the absence of an atom the distributions are Poissonian, whereas in the presence of an atom they are shifted down, following the theoretically-predicted distribution (solid lines). The shaded areas represent the uncertainty in the retrieved distributions due to shot noise.

To establish that the reflected pulse is indeed close to a single-photon Fock state, rather than a classical state of mean 1, we show in Fig. 2b its normalized two-photon detection probability. As evident, for all measured input photon numbers, the statistics are sub-Poissonian, remaining well below the classical limit of 1. Note that Fig. 2b presents the second-order photon-number statistics integrated over the entire pulse duration. It is therefore a much stronger indication for a single-photon state than zero time-delay antibunching $g^2(0)<1$, which reflects only the inability of a single emitter to scatter two photons *simultaneously*. The main limiting factor of the sub-Poissonian behavior is the 4% overlap of the TM mode with the unwanted circular polarization, which leads to the possibility of an additional reflection after the atom underwent a Raman passage - an event that becomes more likely for larger input photon numbers.

The extraction mechanism is further illustrated by the photon-number distributions of the transmitted pulses (Figs. 2c-e), which are essentially the Poissonian distributions of the input pulses shifted down by one photon times the extraction efficiency, in excellent agreement with theory.

The most direct indication of the destructive interference, which lies at the heart of SPRINT and gives rise to the deterministic nature of the photon-extraction scheme, is the temporal correlation between transmitted and reflected photons. The uncertainty in the arrival times of the incoming photons leads to some spread in the ensemble-averaged shape of both the reflected and the transmitted pulses, and accordingly to an apparent overlap between them (Fig. 3a). Yet when examining the arrival times of photons in each individual run (see Fig. 3b for theory and Fig. 3c for experiment), we find that transmission before reflection is a highly unlikely event. The result is a nonclassical, inseparable distribution. The rare occurrence of events in which transmission precedes reflection is due to failures of the extraction mechanism. In our implementation, these failures mostly result from the presence of detuned atomic excited states belonging to the $F'=1$ manifold (see Methods and ref. 13), and from nonlinearities that emerge at a high input photon flux[17] ($>\Gamma^{-1}$).

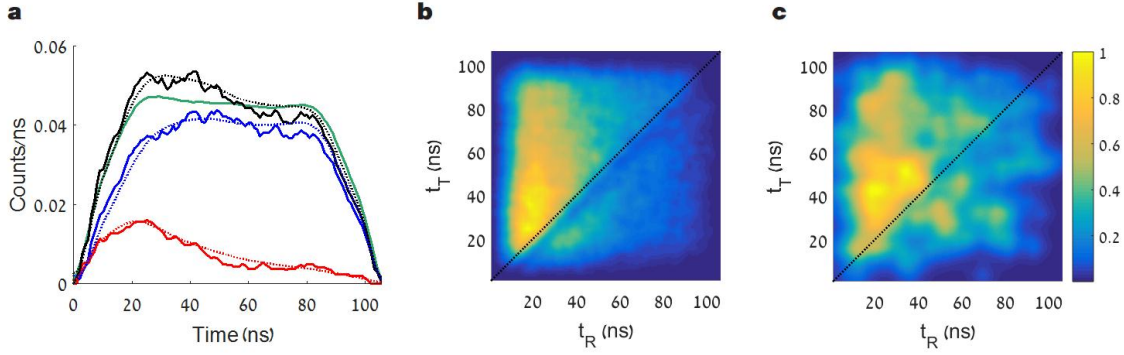

**Figure 3. Nonclassical temporal statistics of the photon-extracted pulse**. **a,** Photon counts as a function of time in the reflected pulse (red), transmitted pulse (blue), and their sum (black) for 11 input photons. The dotted lines show the result of simulations with no free parameters. The measured transmitted pulse in the absence of an atom is shown in green. The overall loss during the SPRINT process is somewhat lower than that of the empty cavity (44% instead of 48%), resulting in slightly higher overall flux (black curve) during this time period. **b, c,** Simulation and measurement, respectively, of the temporal distribution (in arbitrary units) of events in which a single reflected photon and a single transmitted photon were both detected. As expected for a deterministic photon extractor, transmissions are much more likely to occur after a reflection takes place. The data is combined from all measured input intensities.

This demonstration of single-photon extraction is a significant step forward in the ability to control quantum states of light. None of the described limitations are inherent to the scheme, and efficiencies close to unity should be attainable with this system (see Methods and ref. 13) or with other implementations of cavity quantum electrodynamics[29,30]. The extraction mechanism, demonstrated here over a wide range of coherent-state input intensities (from 0.2 photons to 11 photons), is applicable to any input state of light. In particular, a squeezed vacuum at the input would result in an output state close to the highly non-classical Schrödinger kitten state[15]. This versatility, and the fact that the extraction mechanism is insensitive to pulse shape and timing, mark it as a robust building block for the manipulation of light.

**Methods**

**Experimental sequence** Every second a magneto-optically trapped cloud of ~30·10$^6$ atoms at ~7 μK is dropped onto a nanofibre coupled silica microsphere (15 μm radius). When the atomic cloud reaches the microsphere, a sequence of pulses from alternating directions is sent through the nanofibre, and the data acquisition starts. The output photons are separated from the input channel using 95:5 beam splitters, after which they are detected by ten single photon counting modules, five on the reflection port and five on the transmission port. The pulse sequence consists of two detection pulses (the second of which used also as a preparation pulse that initializes the atom in the correct ground state), an 85 ns long measurement pulse of varying strength, an erasure pulse that resets the atomic state in order to remove any detection bias, and two additional detection pulses. All pulses except the measurement pulse contain ~1.5 photons and are ~15 ns long. In order to reduce the amount of false atomic detections to below 1.5%, we require the detection of reflected photons in at least three different pulses per sequence, one of which being the preparation pulse, and one after the measurement pulse, to ensure that the atom was present during the entire sequence. This sequence was repeated ~80,000 times in this experiment.

**Data Analysis** Afterpulsing effects of the single photon counting modules were minimized by omitting any detections recorded within 60 ns of a previous detection by the same module. Consequently, even with five detectors at each output, detection efficiencies temporarily drop with each recorded photon. These effects were taken into account in the experimental data presented in Fig.2, and in the simulations presented in Figs. 3 and 4. Additionally, due to optical losses (~60% transmission) and detector efficiencies (~50%), only ⅓ of the photons are successfully detected upon exiting the setup. The experimental results presented in Figs. 2 and 3a were corrected to account for these losses.

**Retrieval of photon-number distribution**

In order to measure the photon number distribution of the transmitted and reflected pulses, we placed five balanced detectors on either port. In addition to the detector efficiencies and losses mentioned above, one has to take into account that different photons could end up in the same (non photon-number resolving) detector, and will therefore result in only a single click. Given $k$ photons at the input, the probability of detecting $n$ clicks is given by

$$P(n|k) = \frac{(N-N_d)!}{(N-N_d-n)!} \sum_{i=0}^{n} \frac{\left[1 - \eta\left(1 - \frac{i+N_d}{N}\right)\right]^k}{i!\,(n-i)!\,(-1)^{n-i}} \quad, \quad n \leq (N-N_d)$$

where $\eta$ (~⅓ in our case) denotes the total detection efficiency, $N$ the number of detectors, and $N_d$ allows for the possibility that some detectors may be inactive ("dead") already at the beginning of the pulse due to detection events from previous pulses. $P(n|k)$ is then used to construct a matrix $\mathbf{A}$ that transforms the initial distribution $P^{ini}$ into the measured one $P^{meas}$, i.e. $P^{meas} = \mathbf{A}P^{ini}$. This matrix is not square and therefore cannot be inverted in order to retrieve the initial distribution. Even if one neglects the possibility of the presence of more photons than the number of active detectors, so that the matrix could be inverted, the obtained solution is highly sensitive to noise. A better approach is to use a Maximum Entropy regularization algorithm in which we minimize the functional $\Phi[x] = \|\mathbf{A}x - P_{meas}\|^2 + \lambda^2 \sum x \ln x$, subject to the constraint $\sum x = 1$ and enforcing the reconstructed mean to equal the measured one, i.e. $\Sigma_n n \cdot \mathbf{A}x = \Sigma_n n \cdot P_{meas}(n)$.

The regularization parameter $\lambda$ is chosen to be the largest possible one that results in a solution that agrees with the measured values to within a standard deviation, assuming a shot-noise limited measurement.

**Simulations** In order to characterize the behavior of the single-photon extractor theoretically, we first simulate the trajectory of a Rubidium atom in the vicinity of the microsphere, in the presence of Van der Waals forces and gravity. Using analytical solutions for TM whispering gallery modes, we then combine Schrödinger's equation with the input-output formalism[13,17], to obtain the reflection and transmission statistics of a series of single photon pulses interacting with the microsphere and the atom, taking into account losses, pulse shape, multilevel effects, Van der Waals level shifts, magnetic fields, and Rayleigh scattering between counter-propagating modes[13]. We then create a three-dimensional map of the likelihood of the atom being detected at any location. The resulting probability density function is then used to determine the average values mentioned in the text, for example the theoretically derived average value of the atom-photon coupling strength of 24 MHz, and the average polarization properties of the TM mode seen by the atom. Monte Carlo simulations with coherent pulse driving, taking into account the obtained values from the trajectory simulation, optical losses, detector efficiencies and dead times, are then used for obtaining the simulation data presented in the figures. Instances in which the atom ended up in the *m=0* Zeeman ground state were discarded, since the atom loses its coupling to the cavity in that state, and therefore could not be re-detected after the measurement pulse, as required in the actual experiment.

**Experimental limitations**

Loss: The main limitation in the current realization of photon extraction is photon loss. Assuming this is the only imperfection in the system, a theoretical analysis yields the following steady-state solutions: when the atom is in ground state $\beta$, it is transparent to incoming photons. The amplitude transmission coefficient $t_0 = -\frac{\kappa_{ex} - \kappa_i}{\kappa_{ex} + \kappa_i}$ is limited by the microsphere intrinsic loss rate $\kappa_i$, and no light is reflected ($r_0 = 0$). However, when the atom is in ground state $\alpha$, it interacts with incoming photons, and the reflection coefficient for low input photon flux becomes $r = \frac{\kappa_{ex}}{\kappa_{ex} + \kappa_i} \frac{4C}{4C+1}$ (with the cooperativity defined as $4C = \frac{2g^2}{(\kappa_i + \kappa_{ex})\gamma}$), whereas the transmission coefficient becomes $t = t_0 + r$.

The efficiency of SPRINT, defined as $R = |r|^2$, is maximized by tuning the coupling rate between the cavity and the nanofibre (by varying their relative position) to $\kappa_{ex} = \kappa_i \sqrt{1 + \frac{2g^2}{\kappa_i \gamma}}$. This choice of $\kappa_{ex}$ enables maintaining the destructive interference in the forward direction even in the presence of loss, and maximizes the reflection coefficient, which becomes $r = -t_0 \cong \frac{\sqrt{2}g - \sqrt{\kappa_i \gamma}}{\sqrt{2}g + \sqrt{\kappa_i \gamma}}$.

Improving the quality of the microspheres is therefore a promising route for achieving higher efficiencies. For example, decreasing $\kappa_i$ from 6.6 MHz to 0.5 MHz (as we now routinely obtain using improved fabrication procedures), should reduce the losses to ~13% (~25% taking into account also the polarization impurity and the multilevel structure of $^{87}$Rb). Further improvement could be obtained by decreasing the microsphere radius and hence the whispering gallery mode volume. This will increase the atom-photon coupling rate *g*, but might also come at the cost of increased polarization mismatch effects.

Variations in the coupling strength: In practice, one has to choose a value for $\kappa_{ex}$ that maximizes the extraction efficiency for a single value of *g*. However, since in the current implementation the atom is not trapped at a fixed location near the microsphere, the distribution of *g* in our experiment has an approximately Gaussian profile with average 24 MHz and a standard deviation of 9 MHz (as indicated both by the experimental results and numerical simulations). Nevertheless, since the reflection coefficient is proportional to $\frac{4C}{4C+1}$, the efficiency barely depends on *g* for *4C>>1* (as is the case for the majority of detected atoms). Accordingly, the effect of this spread is very limited, and most atoms perform as photon extractors with similar efficiencies.

Polarization impurity: The imperfect circular polarization of the TM mode leads to some undesired coupling of the atom in *β* (*α*) to photons in mode *a* (*b*). This effect can be minimized by using an external magnetic field to induce Zeeman splitting between the ground states, and thus reinforce the directionality. Alternatively, the polarization impurity can be nearly eliminated by using a WGM resonator with higher index of refraction. The overlap with undesired polarization can be estimated by $\frac{1}{2} - \frac{n\sqrt{n^2-1}}{2n^2-1}$, indicating that by switching from silica (*n*=1.45) to, for example, silicon nitride (*n*=2), the polarization impurity is reduced by a factor of 5, from 2.5% to 0.5%. The exact solution of the WGM predicts a similar reduction from 4% to 1%.

Multilevel structure of $^{87}$Rb: Finally, the close presence of another excited state ($F'=1$) has a significant effect on SPRINT. Although the scheme would work as expected with any of the 3-level Λ-systems separately, the steady-state solution of the combined 4-level system can lead to a different result. If the product of the Clebsch-Gordan coefficients of the two transitions has equal signs for both Λ-systems, the symmetry between the output modes is preserved and the contributions of the two systems add up constructively. Yet if this product has opposite signs (as is in our case of $F'=0$ and $F'=1$ ), the symmetry required for SPRINT is broken, leading to degradation in its efficiency. In the extreme case where the two excited states are nearly degenerate, or if the cavity and the probe are tuned halfway between them, the result is complete destructive interference between the probability amplitudes of SPRINT through either excited state.

If the cavity and the probe are resonant with one of the transitions, the dynamics are more complicated but still end up in a reduced SPRINT efficiency as compared to just one excited state (~9% in our case).

Fortunately, by slightly detuning the cavity from atomic resonance, the phase of the combined emission from both Λ-systems can be tuned to restore the destructive interference with the probe in transmission, thereby significantly reducing this degradation (to just 2% in our case, for 20 MHz detuning). A detailed discussion on this issue and all the other limitations and their potential solutions can be found in ref. 13.

based on spin-orbit interaction of light. *Science* **346,** 67–71 (2014).

**Acknowledgments** Support from the Israeli Science Foundation, the Joseph and Celia Reskin Career Development Chair in Physics, and the Crown Photonics Center is acknowledged. This research was made possible in part by the historic generosity of the Harold Perlman family.

**Author Contributions**  All authors contributed to the design, construction and carrying out of the experiment, discussed the results and commented on the manuscript. S.R., O.B. and I.S. analyzed the data and performed the simulations. S.R., O.B. and B.D. wrote the manuscript. S.R. and O.B. contributed equally to this work.

**Author Information** Reprints and permissions information is available at www.nature.com/reprints. The authors declare no competing financial interests. Correspondence and requests for materials should be addressed to B.D. (barak.dayan@weizmann.ac.il).